\documentclass[12pt]{article}

\title{Semi-Analytical Computation of Acoustic Scattering by Spheroids and Disks}
\author{Ross Adelman, Nail A. Gumerov, and Ramani Duraiswami}
\date{}

\usepackage{amsmath}
\usepackage{amsfonts}
\usepackage{amsthm}
\usepackage[bf,labelsep=period]{caption}
\usepackage[margin=1.0in]{geometry}
\usepackage{microtype}
\usepackage[pdftex]{graphicx}
\usepackage{verbatim}

\begin{document}

\maketitle

\section*{Abstract}

Analytical solutions to acoustic scattering problems involving nonspherical shapes, such as spheroids and disks, have long been known and have many applications.
However, these solutions require special functions that are not easily computable.
For this reason, their asymptotic forms are typically used since they are more readily available.
We explore these solutions and provide computational software for calculating their nonasymptotic forms, which are accurate over a wide range of frequencies and distances.
This software, which runs in MATLAB, computes the solutions to acoustic scattering problems involving spheroids and disks by semi-analytical means, and is freely available from our webpage.

\section{Introduction}

Despite the advent of efficient numerical solvers for many physical problems, analytical methods remain valuable.
Many problems can be simplified to a point where analytical methods can be applied.
These methods can provide insight into the problem and can help researchers when moving to a numerical solver.
Numerical methods, on the other hand, have the advantage of being able to treat problems with arbitrary geometries.
In order to do so, they return numerical approximations.
The numerical error can usually be reduced by increasing the number of modeling elements.
Without some kind of validation, though, there is no way to know whether the numerical solution converges to the correct one.
Analytical methods can provide this validation.

The indirect boundary element method (IBEM) for the Helmholtz equation is a numerical method that requires this kind of validation \cite{gumerov2009}.
The IBEM is capable of treating open, infinitely thin surfaces.
These surfaces are good approximations to those often encountered in practice -- those that are much smaller than a wavelength in one dimension, but span several in the other two.
However, there is only one analytically tractable problem posed on such a surface that has an analytical solution and could be used to validate the IBEM: an acoustic wave scattering off a disk.
The disk is actually the degenerate form of the oblate spheroid, so methods for solving scattering problems involving oblate spheroids can also be applied to the disk.
This application is what lead us to the present problem.
Scattering problems involving oblate spheroids, as well as the closely related prolate spheroids, have been studied for well over a century, and analytical expressions for computing their solutions have been documented extensively in the literature.

One of the earliest papers on the topic was by Lord Rayleigh in 1897 \cite{rayleigh1897}.
As discussed in \cite{bouwkamp1950}, prior to the discovery and use of the spheroidal wave functions, the best solutions to these problems were approximations, either in frequency or distance.
When separated in spheroidal coordinates, solutions to the Helmholtz equation can be written in terms of the spheroidal wave functions \cite{flammer2005}.
Because spheroids can be represented as isosurfaces in these coordinate systems, solutions to acoustic waves scattering off them can be written in terms of these special functions.
The resulting expressions are accurate over a wide range of frequencies and distances.
These expressions were applied to the disk in \cite{spence1948, storruste1948}, and they were validated experimentally in \cite{leitner1949}.
Spheroids of different sizes were studied in the following decades \cite{chertock1961, senior1966a}, and much of this work was organized into an encyclopedic book \cite{bowman1987}, which also includes an extensive bibliography on the topic.
More recently, these expressions were implemented and used to compute the solutions over a wide range of frequnecies and spheroid sizes \cite{barton2003, roumeliotis2007}.
However, despite the immense amount of work that has been done on the topic, there are currently no publicly available libraries that implement these expressions.
Also, in all of these references, the spheroids and disks were assumed to be entirely sound soft or sound hard, but the more general case of Robin boundary conditions was never considered in detail.
This work represents our effort to correct this.

We have developed computational software for calculating the solutions to acoustic scattering problems involving prolate spheroids, oblate spheroids, and disks.
This software is called \verb#scattering# and runs in MATLAB.
We have also developed software for computing the spheroidal wave functions required by \verb#scattering#.
This software is called \verb#spheroidal# and is described in a separate paper \cite{adelman2014}.
Using \verb#spheroidal#, we have precomputed many values of the spheroidal wave functions, which, along with \verb#scattering#, are freely available for download from our webpage \cite{scattering_webpage}.
Together, they can be used to recreate the examples seen in this paper.

\section{Spheroidal Coordinates}

\begin{figure}[t]
	\centering
	\includegraphics[height=1.7in]{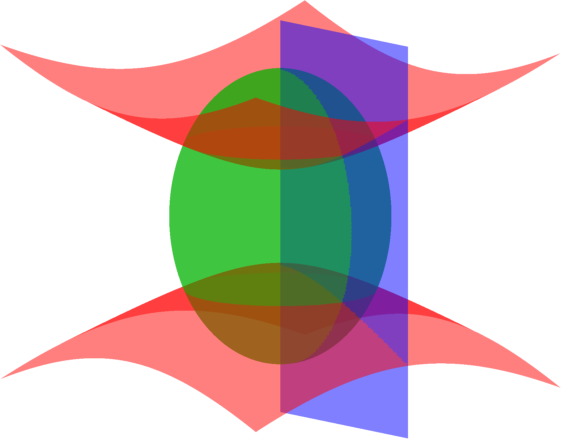}
	\includegraphics[height=1.7in]{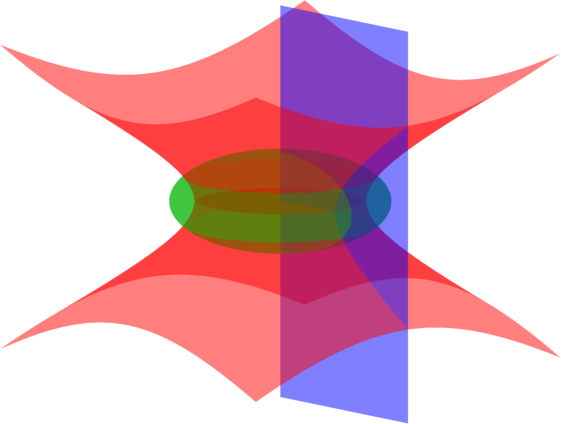}
	\caption{The prolate (left) and oblate (right) spheroidal coordinate systems.  The three colored surfaces are isosurfaces for $\eta = \pm1 / 2$ (red), $\xi = 3 / 2$ for the prolate case and $\xi = 1 / 2$ for the oblate case (green), and $\phi = 0$ (blue).}
	\label{y001}
\end{figure}

The prolate spheroidal coordinate system, shown in Figure \ref{y001}, is related to the Cartesian coordinate system by \cite{flammer2005}
\begin{equation}
x = a\left(1 - \eta^2\right)^{1 / 2}\left(\xi^2 - 1\right)^{1 / 2}\cos\left(\phi\right),\quad
y = a\left(1 - \eta^2\right)^{1 / 2}\left(\xi^2 - 1\right)^{1 / 2}\sin\left(\phi\right),\quad
z = a\eta\xi,
\end{equation}
where $2a$ is the interfocal distance.
The Helmholtz equation, $\nabla^2V + k^2V = 0$, where $k$ is the wavenumber, can be written in prolate spheroidal coordinates as
\begin{equation}
\label{x036}
\left(\frac{\partial}{\partial\eta}\left(1 - \eta^2\right)\frac{\partial}{\partial\eta} + \frac{\partial}{\partial\xi}\left(\xi^2 - 1\right)\frac{\partial}{\partial\xi} + \frac{\xi^2 - \eta^2}{\left(1 - \eta^2\right)\left(\xi^2 - 1\right)}\frac{\partial^2}{\partial\phi^2} + c^2\left(\xi^2 - \eta^2\right)\right)V = 0,
\end{equation}
where $c = ka$.
Applying the method of separation of variables yields three uncoupled ordinary differential equations, one for each coordinate:
\begin{equation}
\label{x004}
\frac{\partial}{\partial{}\eta}\left(\left(1 - \eta^2\right)\frac{\partial}{\partial{}\eta}S_{mn}\left(c, \eta\right)\right) + \left(\lambda_{mn} - c^2\eta^2 - \frac{m^2}{1 - \eta^2}\right)S_{mn}\left(c, \eta\right) = 0,
\end{equation}
\begin{equation}
\label{x017}
\frac{\partial}{\partial{}\xi}\left(\left(\xi^2 - 1\right)\frac{\partial}{\partial{}\xi}R_{mn}\left(c, \xi\right)\right) - \left(\lambda_{mn} - c^2\xi^2 + \frac{m^2}{\xi^2 - 1}\right)R_{mn}\left(c, \xi\right) = 0,
\end{equation}
\begin{equation}
\label{x018}
\frac{\partial^2}{{\partial\phi}^2}\Phi_m\left(\phi\right) + m^2\Phi_m\left(\phi\right) = 0,
\end{equation}
where $m = 0, 1, \ldots$ and $n = m, m + 1, \ldots$.
The solutions to Eq.\ (\ref{x004}) are called the prolate spheroidal angle functions, and the solutions to Eq.\ (\ref{x017}) are called the prolate spheroidal radial functions.
Collectively, they are called the prolate spheroidal wave functions.
Any solution to Eq.\ (\ref{x036}) can be written as
\begin{equation}
\label{pro_sum}
V = \sum_{m = 0}^\infty\sum_{n = m}^\infty{}S_{mn}\left(c, \eta\right)\left(A_{mn}R_{mn}^{\left(1\right)}\left(c, \xi\right) + B_{mn}R_{mn}^{\left(3\right)}\left(c, \xi\right)\right)\cos\left(m\phi\right),
\end{equation}
where the expansion coefficients, $A_{mn}$ and $B_{mn}$, depend on the problem being solved.

The expression for an acoustic wave due to a point source is $V^\text{ps} = \exp\left(ikr\right)/\left(4\pi{}r\right)$, where $r$ is the distance between the point source and evaluation point.
When expanded in terms of the prolate spheroidal wave functions,
\begin{equation}
\label{x022}
V^\text{ps} = \frac{ik}{2\pi}\sum_{m = 0}^\infty\sum_{n = m}^\infty\frac{\epsilon_m}{N_{mn}\left(c\right)}S_{mn}^{\left(1\right)}\left(c, \eta_0\right)S_{mn}^{\left(1\right)}\left(c, \eta\right)R_{mn}^{\left(1\right)}\left(c, \xi_<\right)R_{mn}^{\left(3\right)}\left(c, \xi_>\right)\cos\left(m\phi\right),
\end{equation}
where $\left(\eta_0, \xi_0, 0\right)$ and $\left(\eta, \xi, \phi\right)$ are the positions of the point source and evaluation point in prolate spheroidal coordinates, respectively (because of symmetry, $\phi_0 \neq 0$ can be achieved by rotating the problem around the $z$ axis), $\xi_< = \min\left(\xi_0, \xi\right)$, and $\xi_> = \max\left(\xi_0, \xi\right)$.
Likewise, the expression for a plane wave is $V^\text{pw} = \exp\left(i\bold{k}\cdot\bold{r}\right)$, where $\bold{k}$ is the wavevector and $\bold{r}$ is the position vector of the evaluation point.
When expanded in terms of the prolate spheroidal wave functions,
\begin{equation}
\label{x024}
V^\text{pw} = 2\sum_{m = 0}^\infty\sum_{n = m}^\infty\frac{\epsilon_mi^n}{N_{mn}\left(c\right)}S_{mn}^{\left(1\right)}\left(c, \cos\left(\theta_0\right)\right)S_{mn}^{\left(1\right)}\left(c, \eta\right)R_{mn}^{\left(1\right)}\left(c, \xi\right)\cos\left(m\phi\right),
\end{equation}
where, in this expression, $\bold{k}$ has been restricted to $\bold{k} = k\left(\sin\left(\theta_0\right), 0, \cos\left(\theta_0\right)\right)$ (unrestricted values of $\bold{k}$ can be achieved by rotating the problem around the $z$ axis).

Consider a prolate spheroid, which is described by the isosurface, $\xi = \xi_1$.
The prolate spheroid can be sound soft ($\left[V\right]_{\xi = \xi_1} = 0$), sound hard ($\left[dV/d\bold{n}\right]_{\xi = \xi_1} = 0$), or Robin boundary conditions can be used ($\left[V + \alpha{}dV/d\bold{n}\right]_{\xi = \xi_1} = 0$ and $\alpha = \text{constant}$).
The Robin case lies somewhere between the other two: the prolate spheroid is sound soft when $\alpha = 0$ and sound hard when $\alpha \rightarrow \infty$.
Suppose an incident field, $V^\text{i}$, due to either a point source or plane wave, is generated, which strikes the prolate spheroid.
We wish to compute the scattered field, $V^\text{s}$, from the incident field bouncing off the prolate spheroid.
Depending on the boundary conditions and whether the incident field is due to a point source or plane wave, the exact method of solution is slightly different, but they all follow the same procedure:
(1) the incident field is expanded in terms of the prolate spheroidal wave functions using either Eq.\ (\ref{x022}) or (\ref{x024});
(2) the same is done for the scattered field using Eq. (\ref{pro_sum});
(3) the two expressions are added together to form an expression for the total field in terms of the prolate spheroidal wave functions;
(4) the total field and normal derivative are evaluated at the boundary; and
(5) by using the orthogonality of the prolate spheroidal wave functions, the resulting expression is used to determine the expansion coefficients, $A_{mn}$ and $B_{mn}$, so that the boundary conditions are satisfied.
For the Robin case and a point source,
\begin{equation}
\label{x025}
A_{mn} = 0,\quad{}B_{mn} = -\frac{ik}{2\pi}\frac{\epsilon_m}{N_{mn}\left(c\right)}S_{mn}^{\left(1\right)}\left(c, \eta_0\right)R_{mn}^{\left(3\right)}\left(c, \xi_0\right)\frac{R_{mn}^{\left(1\right)}\left(c, \xi_1\right) + \alpha{}R_{mn}^{\left(1\right)\prime}\left(c, \xi_1\right)}{R_{mn}^{\left(3\right)}\left(c, \xi_1\right) + \alpha{}R_{mn}^{\left(3\right)\prime}\left(c, \xi_1\right)}.
\end{equation}
For the Robin case and a plane wave,
\begin{equation}
\label{x026}
A_{mn} = 0,\quad{}B_{mn} = -2\frac{\epsilon_mi^n}{N_{mn}\left(c\right)}S_{mn}\left(c, \cos\left(\theta_0\right)\right)\frac{R_{mn}^{\left(1\right)}\left(c, \xi_1\right) + \alpha{}R_{mn}^{\left(1\right)\prime}\left(c, \xi_1\right)}{R_{mn}^{\left(3\right)}\left(c, \xi_1\right) + \alpha{}R_{mn}^{\left(3\right)\prime}\left(c, \xi_1\right)}.
\end{equation}
For a prolate spheroid that is sound soft ($\alpha = 0$) or sound hard ($\alpha \rightarrow \infty$), these expressions reduce to those in \cite{bowman1987}.

The expressions arising in the oblate case are very similar to (and sometimes exactly the same as) those arising in the prolate case.
In many cases, simply letting $c, \xi \rightarrow -ic, i\xi$ provides a transformation from the prolate case to the oblate case \cite{flammer2005}.
Indeed, the preceeding paragraphs and equations for the prolate case can be transformed into those for the oblate case by using this transformation.
The oblate spheroidal coordinate system is shown in Figure \ref{y001}.

\section{Scattering Routines}

\begin{figure}[t]
	\centering
	\includegraphics[height=1.7in]{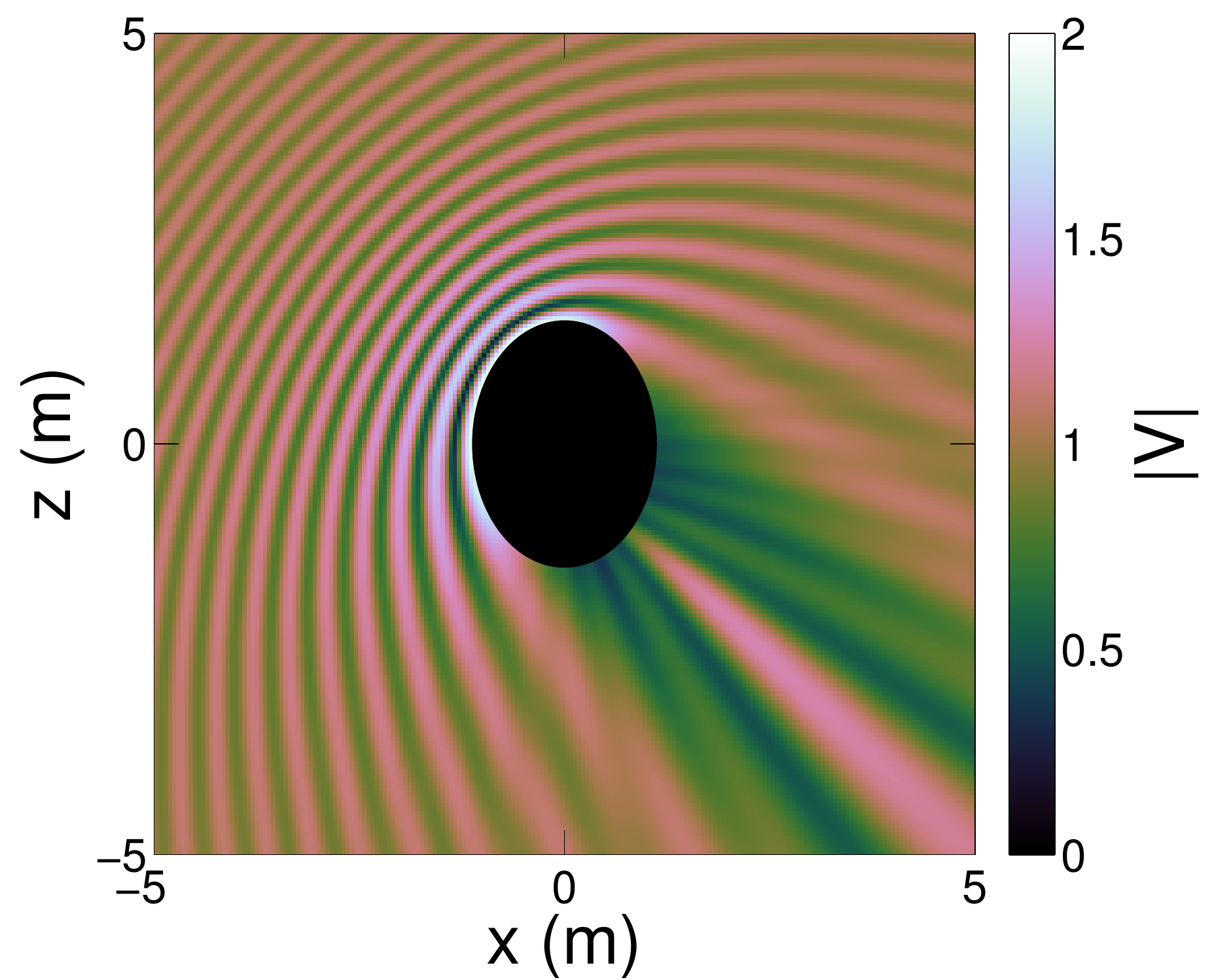}
	\includegraphics[height=1.7in]{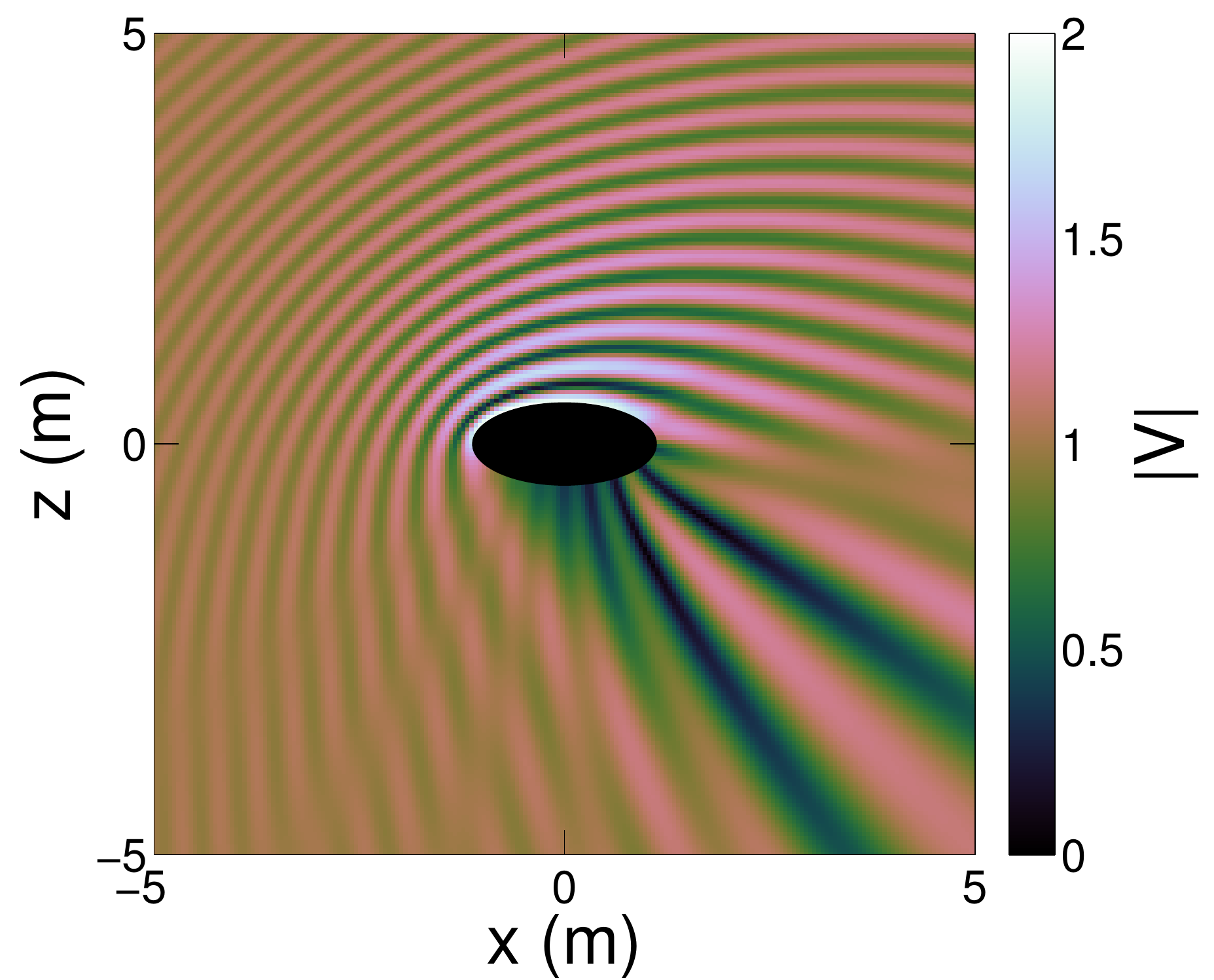}
	\includegraphics[height=1.7in]{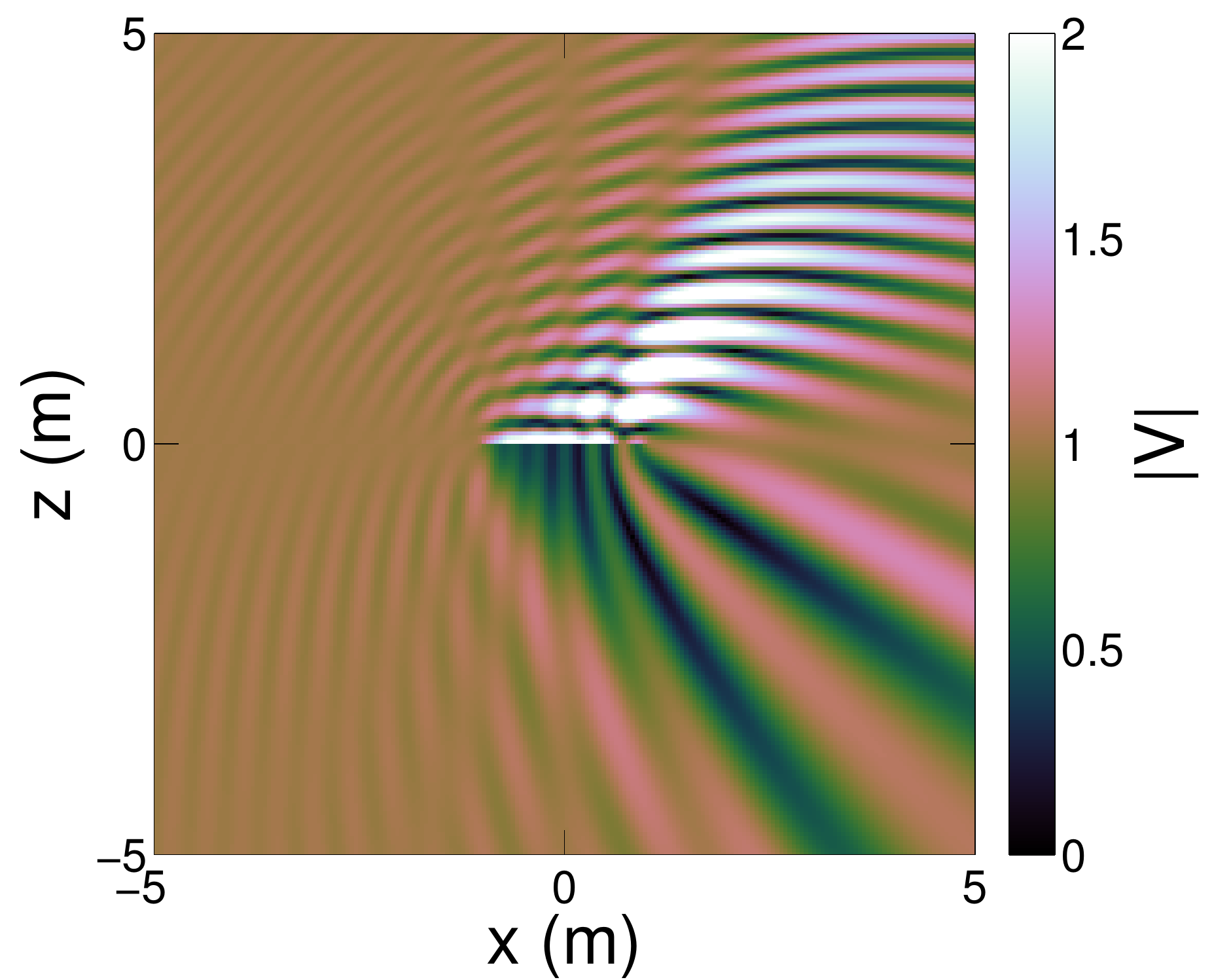}\\
	\includegraphics[height=1.7in]{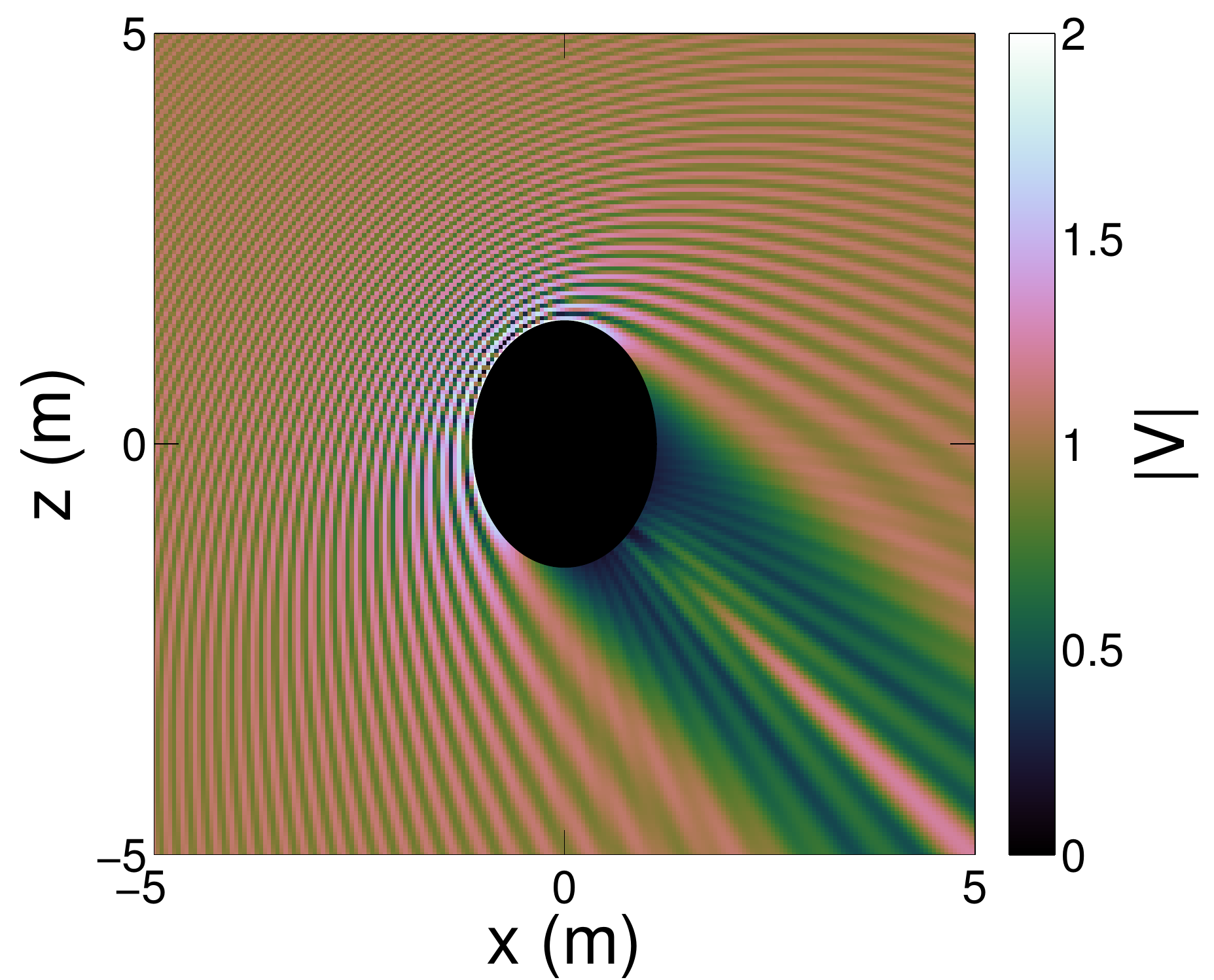}
	\includegraphics[height=1.7in]{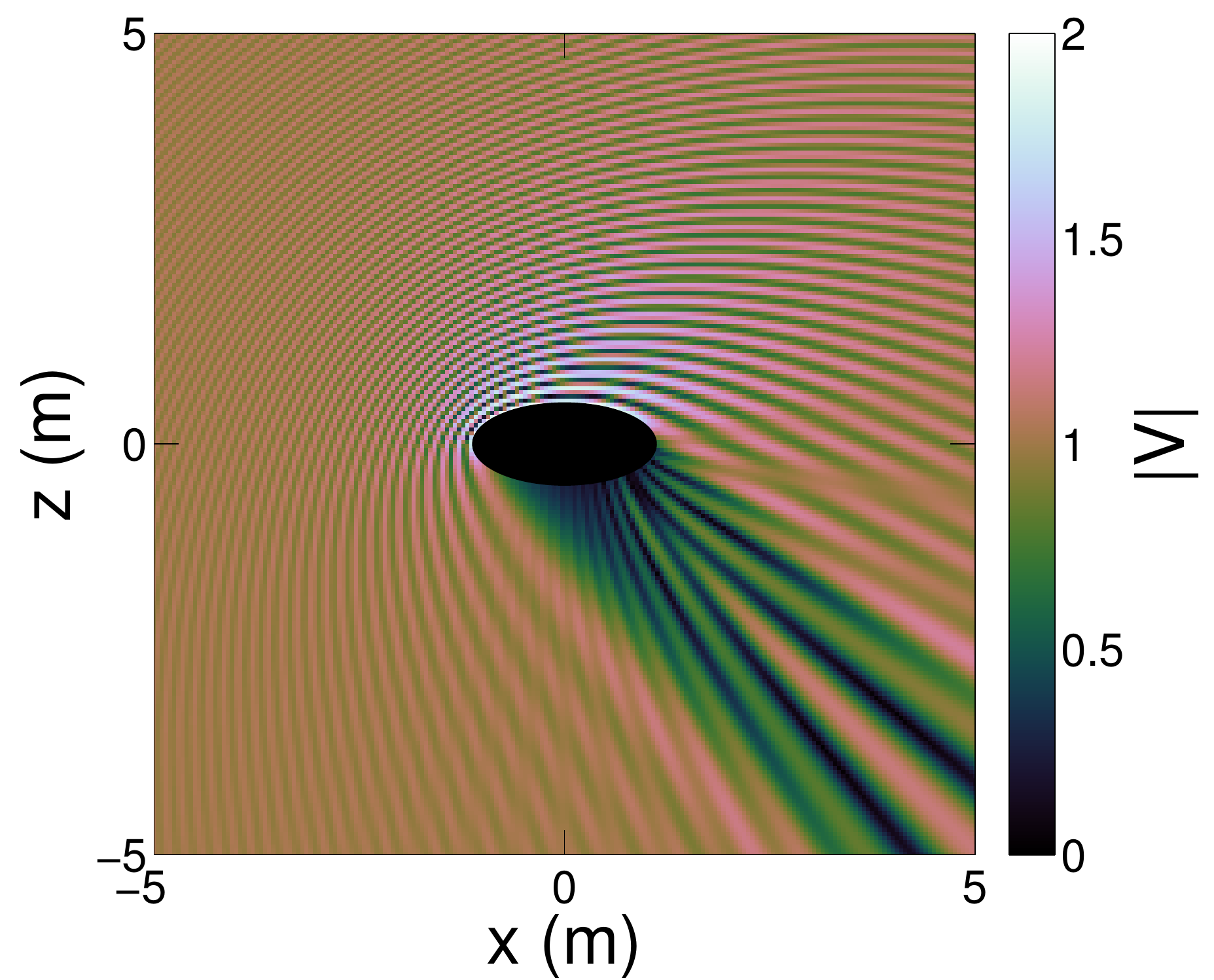}
	\includegraphics[height=1.7in]{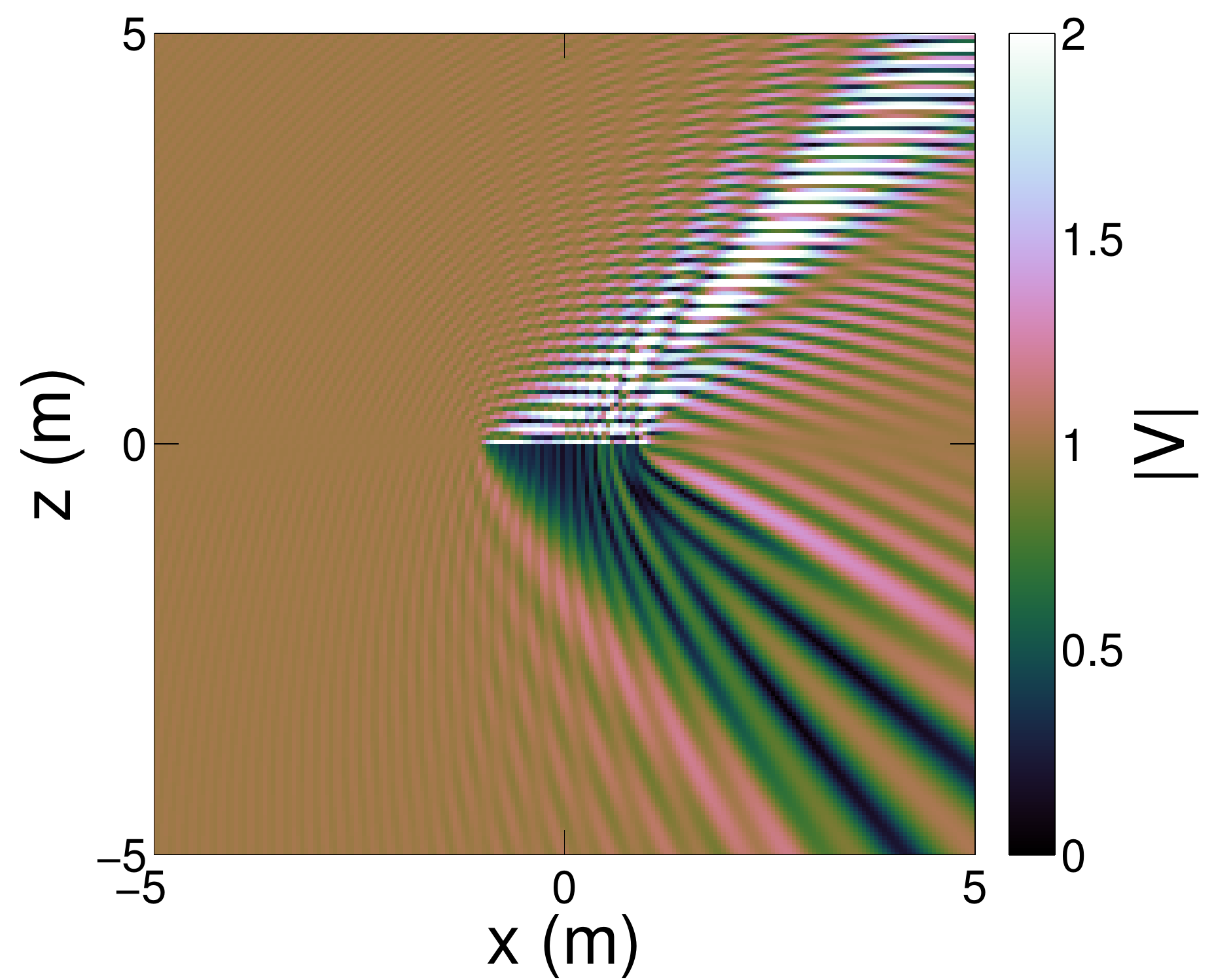}
	\caption{A plane wave scattering off a sound-hard prolate spheroid (left), oblate spheroid (center), and disk (right) for $k = 10$ (top) and $k = 25$ (bottom).}
	\label{y007}
\end{figure}

\begin{figure}[t]
	\centering
	\includegraphics[height=1.7in]{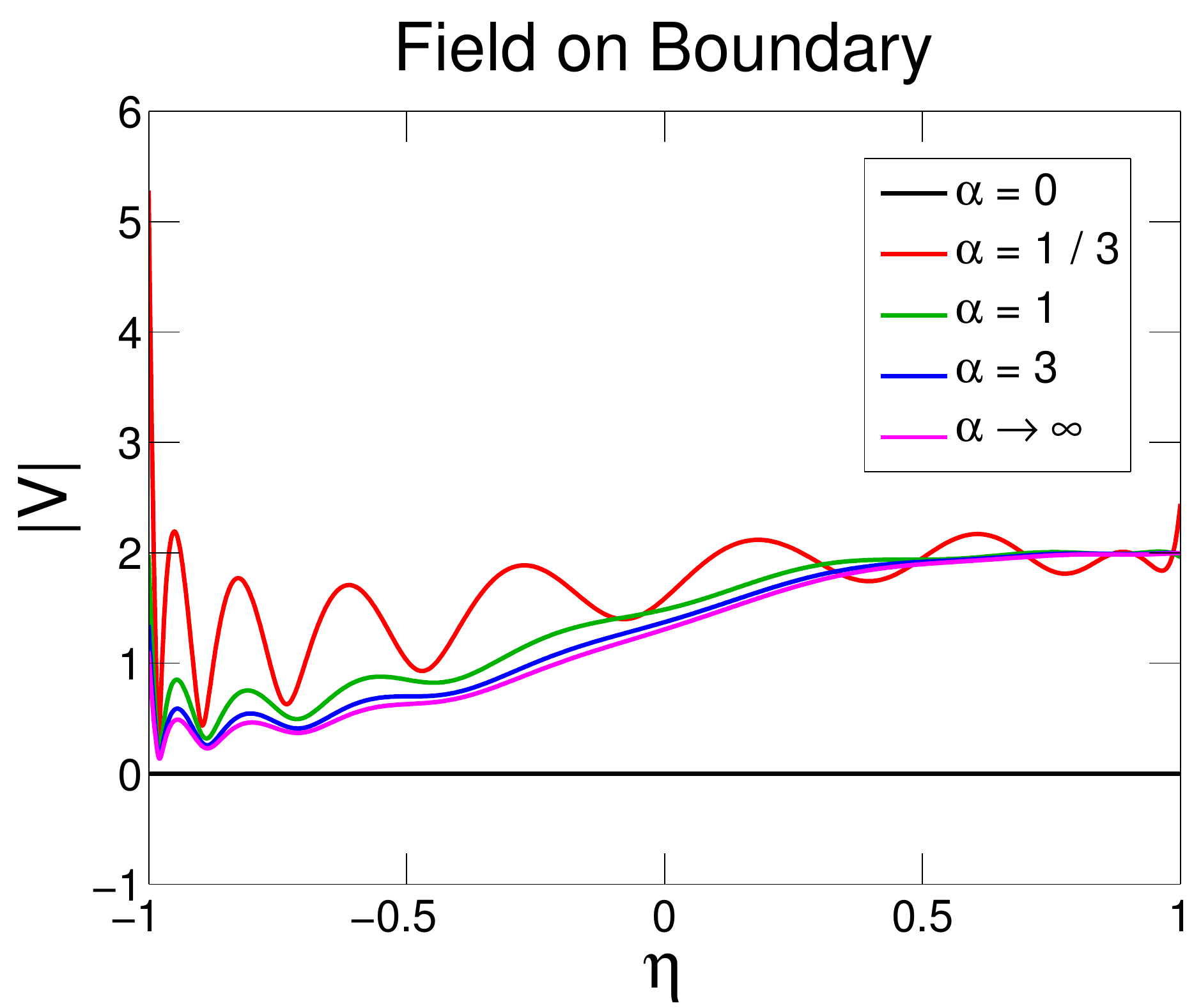}
	\includegraphics[height=1.7in]{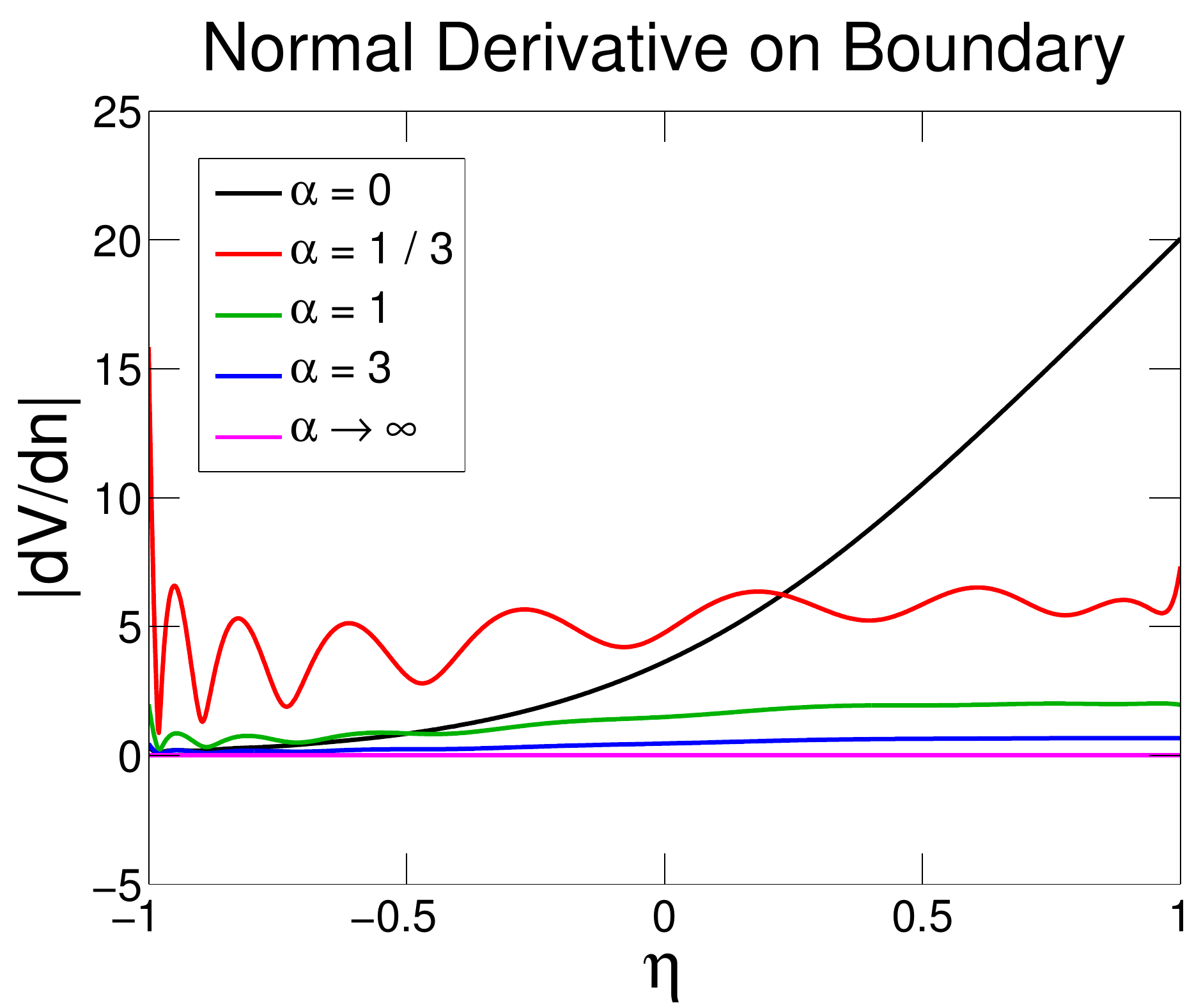}
	\includegraphics[height=1.7in]{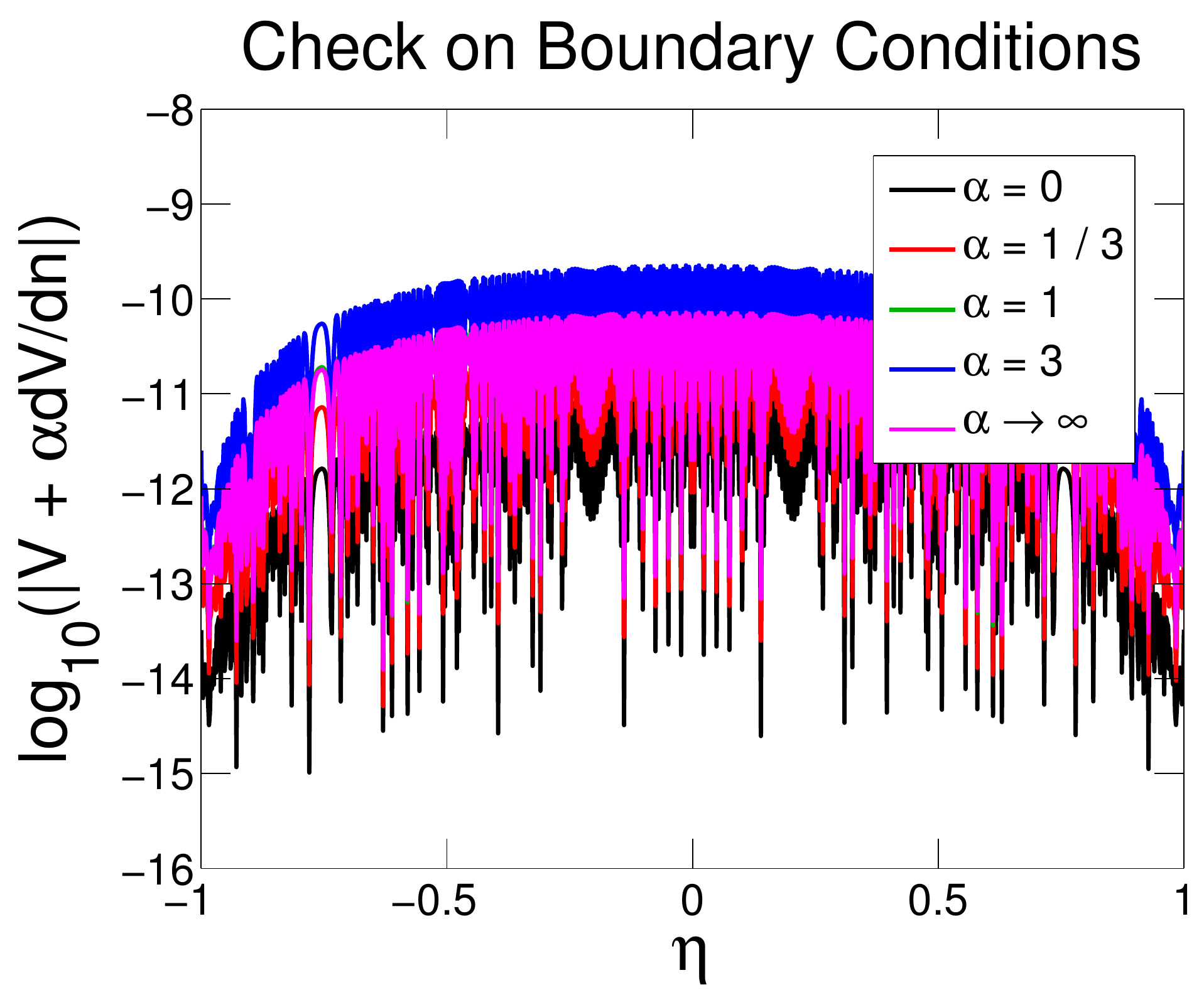}
	\caption{The two figures on the left show the field and normal derivative along the boundary of an oblate spheroid from a plane wave striking the oblate spheroid from directly above for five different choices of the boundary conditions.  The one figure on the right shows that the boundary conditions have been satisfied in all five cases.}
	\label{y010}
\end{figure}

We have developed computational software for calculating the solutions to acoustic scattering problems involving prolate spheroids, oblate spheroids, and disks.
This software is called \verb#scattering# and runs in MATLAB.
The spheroids and disks can be sound soft, sound hard, or Robin boundary conditions can be used, and the incident field can be due to either a point source or plane wave.
Internally, \verb#scattering# needs to compute the spheroidal wave functions of different order and degree for different values of their argument.
We have developed software for doing so.
This software is called \verb#spheroidal# and is described in a separate paper \cite{adelman2014}.

There are 18 routines in \verb#scattering#, each one solving a different scattering problem.
For example, one is called \verb#pro_plane_wave_scat_hard#, which computes the scattered field from a plane wave striking a sound-hard prolate spheroid.
They are organized into 18 separate MATLAB M-Files, one for each routine.
At the top of each M-File is a detailed explanation of the calling convention, return values, and so on.

Each routine has a slightly different calling convention depending on what is being computed, but they all follow the same pattern.
They all require $k$ and $a$, as well as information about the incident field, either $\eta_0$ and $\xi_0$ for a point source or $\theta_0$ for a plane wave.
Routines that compute a scattered field also require $\xi_1$.
For example, to compute the scattered field from a plane wave striking a sound-hard prolate spheroid, the following MATLAB code fragment can be used:
\begin{verbatim}
v_scat = pro_plane_wave_scat_hard(10.0, 1.0, pi, 'saved', 1.5, x, y, z);
\end{verbatim}
where, in this case, $k = 10$, $a = 1$, $\theta_0 = \pi$, \verb#'saved'# is the directory in which the spheroidal wave functions have been precomputed and saved, and $\xi_1 = 3 / 2$.
The variables, \verb#x#, \verb#y#, and \verb#z#, are row vectors, which contain the positions of the evaluation points in Cartesian coordinates, and \verb#v_scat# is a row vector, which will contain $V^\text{s}$.

\section{Conclusion}

We have developed computational software for calculating the solutions to acoustic scattering problems involving prolate spheroids, oblate spheroids, and disks.
While these problems have been studied for well over a century and analytical expressions for computing their solutions have been documented by many authors, there are currently no publicly available libraries that implement them.
This paper gave an overview of these problems, derived their analytical solutions, described some theory behind the special functions required by them, and included several examples of running our software.
We hope that our software will better enable research involving spheroids and disks.

\bibliographystyle{jasanum}
\bibliography{scattering}

\end{document}